\documentclass[twocolumn,superscriptaddress]{revtex4}

\usepackage{amsmath}  
\usepackage{amsfonts} 
\usepackage{graphicx} 

\usepackage{color}

\begin{document}


\title{Revisiting Aristarchus's method to estimate the ratio of the distances from Earth to the Sun
and the Moon}

\author{Hugo Caerols}
\email{hugo.caerols@uai.cl}
\affiliation{Facultad de Ingenier\'{\i}a y Ciencias, Universidad Adolfo Ib\'a\~nez, Santiago 7941169, Chile.}

\author{Mart\'{\i}n Avenda\~{n}o}
\email{mavend01@ucm.es}
\affiliation{Departamento de \'Algebra, Geometr\'{\i}a y Topolog\'{\i}a, Facultad de Matem\'aticas,
Universidad Complutense de Madrid, España.}

\author{Felipe A. Asenjo}
\email{felipe.asenjo@uai.cl}
\affiliation{Facultad de Ingenier\'{\i}a y Ciencias, Universidad Adolfo Ib\'a\~nez, Santiago 7941169, Chile.}

\author{Rafael Brahm} 
\email{rafael.brahm@uai.cl }
\affiliation{Facultad de Ingenier\'{\i}a y Ciencias, Universidad Adolfo Ib\'a\~nez, Santiago 7941169, Chile.}


\date{\today}

\begin{abstract}
In this article we revisit Aristarchus’ ancient experiment to determine the ratios between the Earth–Moon and Earth–Sun distances. We present here the construction of  a simple instrument, basically a tube that, by projecting its shadow, that allows one to estimate the Sun–Earth–Moon angle at the moment of first (or last) quarter. The simplicity of its construction and the accuracy of the measurements obtained may come as a surprise. We discuss the complexities involved in determining the exact moment of first quarter, based on the shape of the elliptical shadow projected on the Moon. Finally, we analyze and implement a strategy that allows us to approximate the measurement of this angle and estimate its value using observations that can be performed twice a month, and more accurately a couple of times during the year. We place Aristarchus’ problem in its historical context and detail the physical and mathematical topics that can be developed and motivated through this experiment, in addition to carrying out a very thorough analysis of the errors associated with this measurement.
\end{abstract}

\maketitle 


\section{\label{sec-intro}Introduction}

Determining the dimensions of the known universe has always held a particular charm, perhaps because
of the natural human curiosity to explore the world in which they are immersed. One of the first to decisively
confront these questions were the Greeks. They were well-versed in Geometry, thanks to the schools of 
Pythagoras of Samos and Thales of Miletus, and mastered the concepts of proportions and similarities of 
figures~\cite{Eves69}. More than that, they were equipped with logical and deductive reasoning, so 
prominently featured in Euclid's Elements~\cite{Euclid}, which remains the legacy of Greek mathematicians 
and philosophers to our generation, who can only marvel at their great discoveries.

Aristarchus was a Greek physicist and mathematician (c.~310--230 BC) who undertook the enormous task
of determining the dimensions of the universe~\cite{Aris59}. He  was the first known astronomer to propose
a heliocentric model. His estimates of the relative sizes of the Moon, Earth, and Sun showed the Sun to be 
substantially larger than the Earth, a result that may have been one of the considerations that led him to place
the Sun rather than the Earth at the center of the universe \cite{Heath}. It is well known that he established
the ratio $\lambda$ between the distance from the Earth to the Sun and from the Earth to the Moon by
estimating the angle $\delta$ between these two celestial bodies when the Moon is in its first quarter,
forming a right triangle with the Sun \cite{Polya77, Ballesteros}. His estimation for this angle was
$87^\circ$, which gives a ratio $\lambda=\sec(\delta)\approx 19$.

\begin{figure}
\includegraphics[width=8cm]{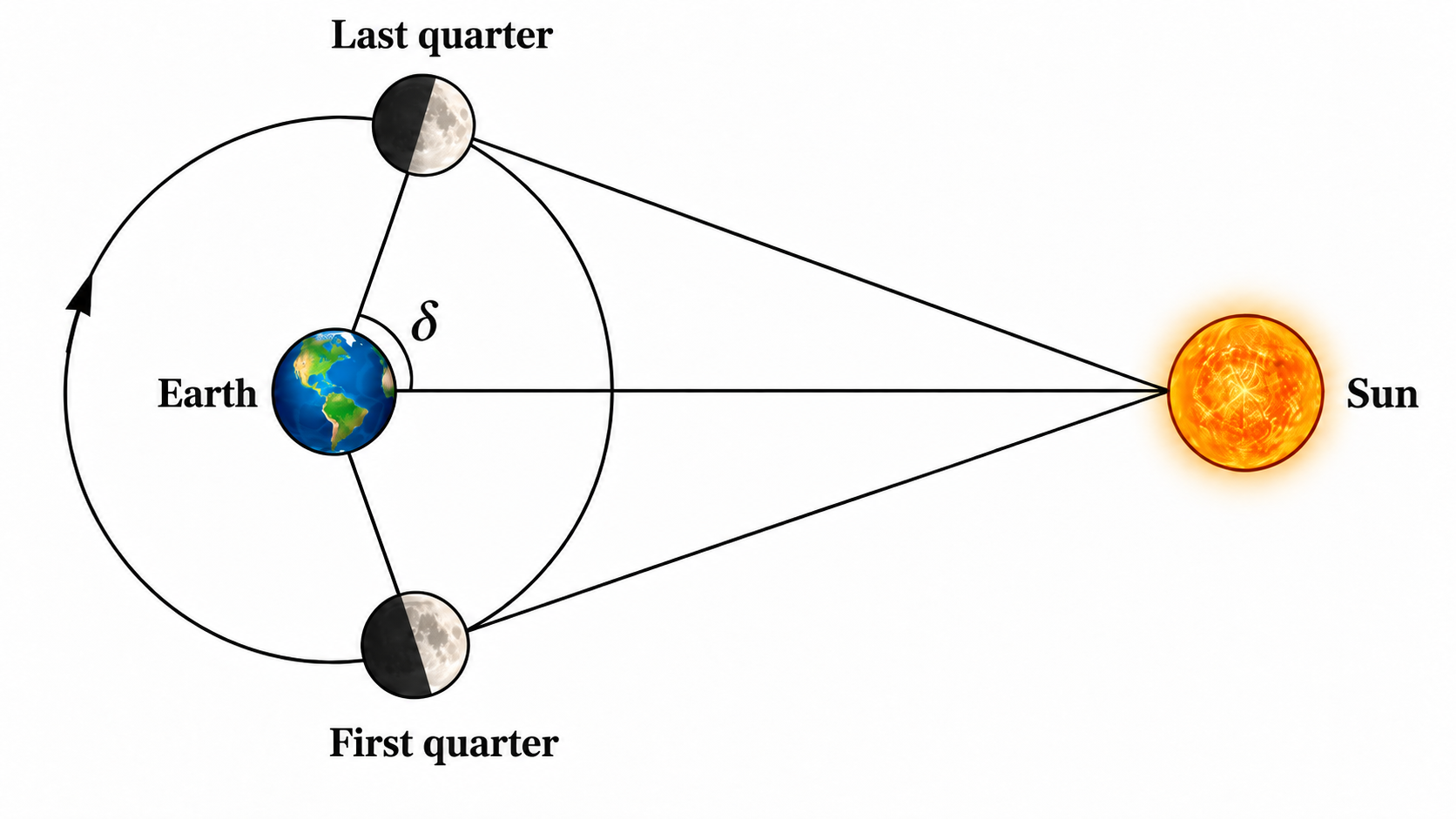}    
\caption{\label{Maza}Positions of the Sun, the Earth and Moon in its first and final quarters.}
\end{figure}

However, exactly how he arrived at this value remains a mystery. We have only found indirect ideas about 
this calculation. In a recently written book about the Moon~\cite{Maza}, the author suggest a reasoning
method leading to this approximate measurement. Aristarchus could have estimated a difference of one day
between the time from a first quarter to a last quarter Moon compared to the time from a last quarter
to a first quarter Moon (see FIG.~\ref{Maza}).  In this way, if the synodic period of the Moon
is estimated at $29$ days, this  angle would correspond to $\delta\approx (14/29)\cdot (360^\circ/2)
\approx 87^\circ$. 

Whether the $\delta\approx 87^\circ$ value resulted from a geometric estimation or a measurement is
unclear from Aristarchus's writings. It is very hard to do a direct measurement of the angle, since
observing the Sun directly is not easy and can even be dangerous if done without proper precautions.
The technical challenges involved in the measurement, as well as the precision required in order to
get an accurate value of the ratio $\lambda$, makes revisiting Aristarchus's ideas a worthy pursuit.

The goal of this work is to present a method for measuring this angle that can be easily replicated
in schools and astronomy academies. To achieve this, we have tried to avoid overly sophisticated
theoretical elements, rediscovering the challenges Aristarchus likely faced and overcoming them using
everyday materials and some logical reasoning. This work also provides excellent methodological
tools for introducing and applying plane geometry and trigonometry~\cite{Cox67} with students. We
hope this will become a replicable activity that inspires and excites those who are willing to take
on Aristarchus's challenge of determining the value of $\lambda$ for themselves.

\begin{figure}
\includegraphics[width=8.6cm]{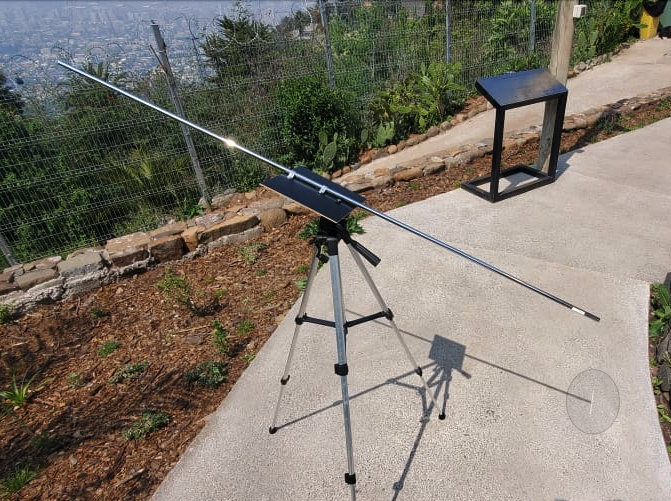}       
\caption{\label{Genial2}The measuring instrument. The shadow of the tube can be seen projected onto
the ground on the terrace of the Foster Observatory on San Cristóbal Hill, Santiago, Chile, where
the experiment was performed. Below, in a circle, we can see the marker indicating that end of
the shadow.}
\end{figure}

To carry out this experiment, we will need to choose a time near the first quarter phase when both the
Sun and the Moon are visible in the sky at the same time. We want to point out that in the measurements 
we will not use telescopes or sophisticated equipment, but a simple instrument, basically a tube that, 
by projecting its shadow, allows us to take measurements and obtain the desired angle (see FIG.~\ref{Genial2}).


 It is astonishing how such an apparently simple experiment can reveal 
sophisticated elements, as will be shown throughout the article.

This paper is organized as follows: In section~\ref{sec-mates}, we present the mathematics behind
Aristarchus's method, discussing the relationship between the angle $\delta$ and the value of
$\lambda$. In section~\ref{sec-aparato}, we describe a very simple device that can be used to measure
the angle $\delta$ with a reasonable accuracy. The idea is to use a $1.5{\rm m}$ thin tube that is
mounted on a tripod, point the tube towards the Moon, and measure the lengths of the trapezoid whose 
vertices are the ends of the tube and their corresponding shadows projected by the Sun on a flat ground 
surface. (see FIG.~\ref{Genial}).
In section~\ref{sec-accuracy}, we study how measurement errors in the lengths of the trapezoid propagate
to the value of $\delta$ by running a simulation of the experiment.
In section~\ref{sec-time}, we discuss the difficulties involved in determining the exact moment of first quarter and present the idea of estimating the angle using photographs of the Moon, which we expect to develop in a separate work due to its extent.
In section~\ref{sec-realdata}, we work with real data measured at the Foster Observatory, Santiago, Chile
on Oct 10th, 2024.
In section~\ref{sec-errors}, we analyze all sources of errors in the experiment, and propose different
ways to mitigate them. Finally, in section~\ref{sec-conclusions}, we summarize the main findings of
in this article, and provide some important tips for anyone who wants to replicate our experiment.


\section{\label{sec-mates}The mathematics behind Aristarchus's method}

\begin{figure}
\includegraphics[width=9cm]{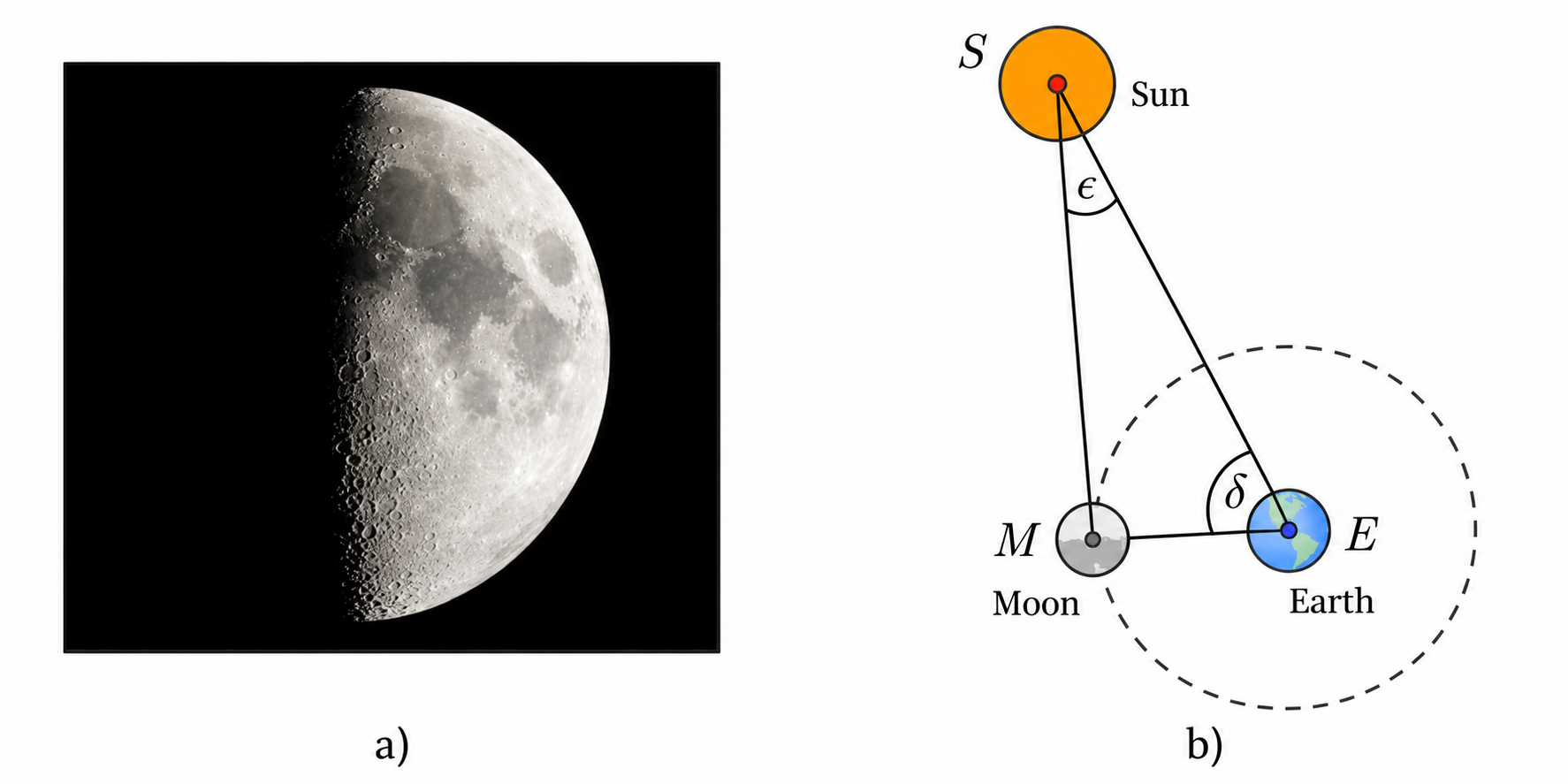}    
\caption{\label{Triangulo2}Part (a) shows a picture of the Moon and part (b) shows the
Earth-Moon-Sun configuration, both at the first quarter.}
\end{figure}

Aristarchus realized some quite simple yet very important facts. The first is that the Sun, the Moon,
and the Earth (the location from which the data is taken) can be thought of as the vertices of a
triangle $SME$. The second is that when the Moon is exactly in its first quarter (or last quarter), the
angle $SME$ is $90^\circ$, i.e. the triangle $SME$ is a right triangle (see FIG.~\ref{Triangulo2}b).

Note that the ratio $\lambda$ between the Earth-Sun distance $ES$ and the Earth-Moon distance $EM$
is directly related to the angle $\delta=SEM$ through the following trigonometric formula:
\[
   \lambda=\frac{ES}{EM}={\rm sec}(\delta).
\]

We mentioned in the introduction that Aristarchus estimated $\delta\approx 87^\circ$. However, a
simple computation will reveal the actual value of this angle.

With the aid of modern astronomical instruments and precise measurements, we now know that the
Sun is about $400$ times farther from the Earth that the Moon, i.e. $\lambda\approx 400$. In
particular, the angle $\varepsilon=ESM$ is very small. Of course, at the half moon, we have
\[
   \delta+\varepsilon=\frac{\pi}{2}
\]
as shown in FIG.~\ref{Triangulo2}, so the problems of estimating $\delta$ or $\varepsilon$ are
equivalent. Since $\varepsilon$ is very small, we can approximate
\[
   \varepsilon\approx\sin(\varepsilon)=\frac{EM}{ES}=\frac{1}{\lambda}\approx\frac{1}{400},
\]
which corresponds to $\varepsilon\approx 9'$. This shows that the actual value of $\delta$
should be approximately $\delta\approx 89^\circ 51'$.

We can reverse the procedure above to obtain an approximate formula for $\lambda$ in terms
of the measure angle $\delta$. Doing so, we get
\[
    \lambda = \frac{1}{\sin(\varepsilon)}\approx\frac{1}{\varepsilon}
            = \frac{1}{\frac{\pi}{2}-\delta},
\]
which is a simpler alternative to $\lambda={\rm sec}(\delta)$ when the angle $\delta$ is close
to $90^\circ$. It is clear from this formula that a small error in $\delta$ can give very different
values of the ratio $\lambda$, as shown in TABLE~\ref{Tabla1}. This highlights the importance of
an accurate value of $\delta$.

\begin{table}
\caption{\label{Tabla1}$\lambda$ as a function of $\delta$.}
\begin{ruledtabular}
\begin{tabular}{cc}
$\delta$ & $\lambda={\rm sec}(\delta)$   \\
\hline
$87^{\circ}$     & $19.11$  \\
$88^{\circ}$     & $28.65$  \\
$89^{\circ}$     & $57.30$  \\
$89^{\circ}30'$  & $114.59$ \\
$89^{\circ}51'$  & $381.97$ \\
$89^{\circ}52'$  & $429.72$ \\
\end{tabular}
\end{ruledtabular}
\end{table}

Assuming a synodic period of $29.5$ days, the angle $\delta$ increases at a rate of $12.2^\circ/
{\rm day}\approx 30.5'/{\rm hour}=30.5''/{\rm min}$. In particular, an error of $2$ minutes in the
time at which the estimate for $\delta$ is made will introduce about $1'$ of error. This is one of
the main technical challenges of Aristarchus's method.
 
 Strictly speaking, the configuration required by Aristarchus's method
is lunar dichotomy, i.e. the instant at which exactly one half of the
lunar disk is illuminated as seen from the observing site. This instant
should be distinguished from the conventional first- and last-quarter
phases listed in astronomical calendars. The two events occur close to
one another, but they are not exactly simultaneous. In what follows,
we therefore use the terms waxing and waning dichotomy for the precise
configurations relevant to the experiment.

\section{\label{sec-aparato}A simple device to measure the angle}

As incredible as it may seem, our measuring instrument only uses simple items such as a curtain rod
and a photographic tripod, both of which are easy to obtain. Instead of aiming at the Sun and the
Moon simultaneously with a device, we will only need to aim the tube at the Moon and work with the
projected shadows to compute the desired angle $\delta$ using trigonometry.

\begin{figure}
\includegraphics[width=8.6cm]{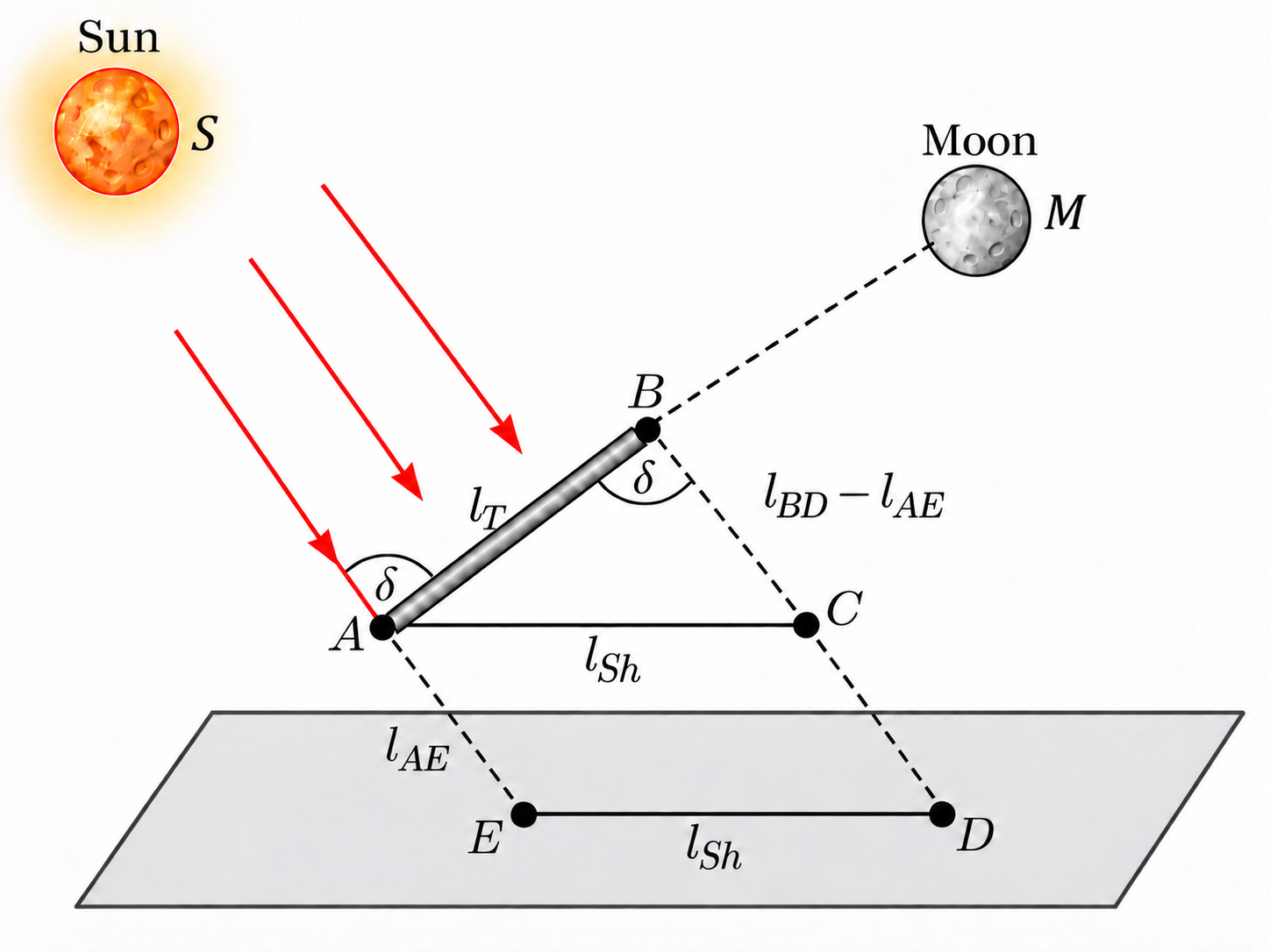}
\caption{\label{Genial}A tube $AB$ of length $l_T$ is aimed at the Moon, casting a shadow $ED$
of length $l_{Sh}$. The lines $AE$ and $BD$ can be regarded as parallel, since the Sun is too far
away from the device.}
\end{figure}

\begin{figure}
\includegraphics[width=7cm]{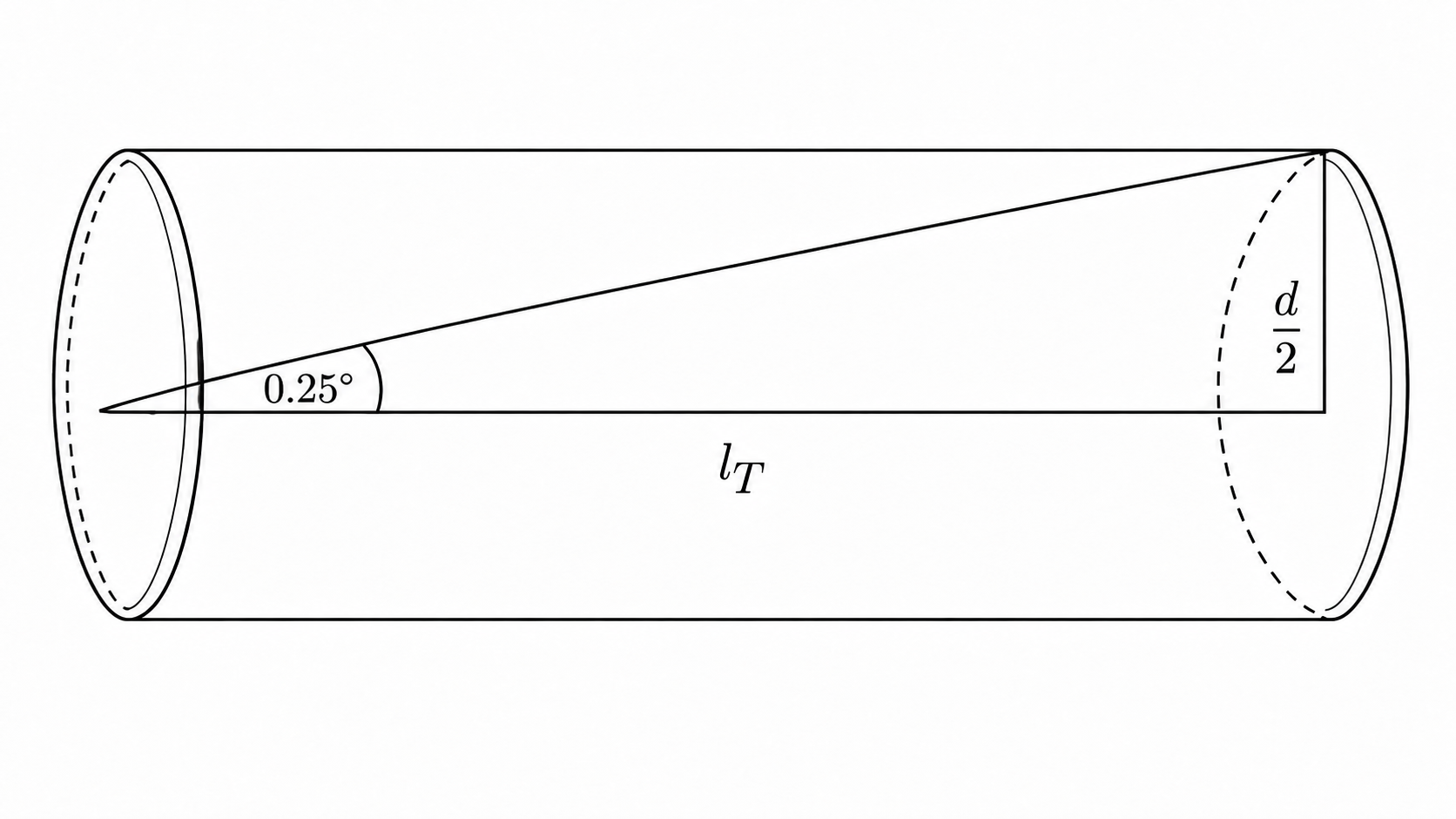}    
\caption{\label{Tubo}Minimum diameter $d$ of a tube of length $l_T$ to be able to see the Moon when
looking through it.}
\end{figure}

A schematic version of our proposed instrument is shown in FIG.~\ref{Genial}. The tube $AB$ should
be of length $l_{T}\approx 1.5{\rm m}$ and made of a material that does not bend too easily. The
tube should be mounted on a tripod that allows for easy aiming and that can be fixed at certain
position when desired. The diameter of the tube should be enough for the Moon to be seen through it.
Since the Moon spans about $0.5^\circ$ on the sky, the minimum diameter of the tube can be computed as
\[
   d = 2 l_{T} \tan(0.25^\circ),
\]
as shown in FIG.~\ref{Tubo}. We suggest a diameter of $d\approx 1.4{\rm cm}$ for a tube of length 
$l_T=1.5{\rm m}$ or $d\approx 1.8{\rm cm}$ for a tube of length $l_T=2.0{\rm m}$. For our experiments
we used a $l_T=1.5{\rm m}$ curtain rod with a diameter of $1.1 {\rm cm}$, just as we found it in
the store, and it has been sufficient to achieve excellent results. We chose to use a tube with a
small diameter and not too long to have greater maneuverability, well-defined shadow edges, and to
reduce errors  associated with these measurements.

The points $A$ and $B$ in FIG.~\ref{Genial} represent the ends of the tube pointing at the Moon. The
segment $ED$ represents the shadow of the tube on the ground, whose length will be denoted as $l_{Sh}$.
Note that the Sun, the Moon, the tube (i.e, the points $S$, $M$, $A$ and $B$), as well as the shadow
(i.e., the points $E$ and $D$), are all coplanar. The imaginary line $AC$ is parallel to the shadow
$ED$. Assuming that the rays of the Sun are parallel due to the small size of the segment $AB$, it
can be deduced that $\delta=\angle ABC$. At this point, we have all the geometry needed to calculate
the  sides of triangle $ABC$. To do this, we just need to measure the distances from each end of the
tube to its  projection on the ground; that is, we need to measure the distance between points $B$
and $D$, which we will call $l_{BD}$, and between $A$ and $E$, which we will call $l_{AE}$. If we
denote $l_{BC}$ as the length of side $BC$, it can be calculated as $l_{BC}=l_{BD}-l_{AE}$, and also 
$l_{AC}=l_{Sh}$, since $ACDE$ is a parallelogram. Finally, by applying the cosine theorem to
$\triangle ABC$, we get
\begin{equation}\label{eqdelta}
   \delta = \arccos\left(\frac{l_{T}^2 + l_{BC}^2 - l_{Sh}^2}{2l_{T}l_{BC}}\right).
\end{equation}
This method computes $\delta$ indirectly from the lengths of the sides of the trapezoid $ABDE$,
but does not require any angle measurement (which are very hard to do with a $1'$ of accuracy)
nor pointing the device to two objects in sky simultaneously. Moreover, the value of $l_T$ is
constant and can be determined in advance as precisely as needed.

Since our calculations of the angle $\delta$ are based on linear measurements, we analyze how the precision of these measurements affects the calculation of the angle $\delta$. To this end, we generate exact data for the measurements corresponding to a given angle $\delta$, and then assume errors of a given magnitude in these data. We subsequently evaluate how far the angle $\delta$, calculated using Eq.~\eqref{eqdelta}, deviates from the exact angle initially assumed.

\section{Expected accuracy of the calculations}
\label{sec-accuracy}


Since we do not know the initial position of the Moon, we introduce as a parameter the angle $\angle BAC=\alpha$ in Fig.~\ref{Genial}, which we use for this analysis. The length of the tube used is $l_{T}=1.5$~[m]. We can assume $AE=1$~[m], since we can always theoretically lower the tube as close to the ground as desired and adjust the setup so that the shadow length is one meter. In triangle $ABC$, note that $\angle ACB=180-(\alpha+\delta)$. Applying the sine theorem to this triangle, we obtain
$$
BC=\dfrac{l_{T}\sin(\alpha)}{\sin(\alpha+\delta)} \quad \text{and} \quad
l_{Sh}=\dfrac{l_{T}\sin(\delta)}{\sin(\alpha+\delta)}.
$$

For example, if $\alpha=30^{\circ}$ and $\delta=87^{\circ}$, the exact measurements at that moment are
$l_{T}=1.5$~[m], $l_{Sh}=1.681182$~[m], $AE=1$~[m], and $BD=1.841745$~[m].\\

If we use an instrument such as a large ruler or a measuring tape with centimeter resolution, our measurements will be approximated to three decimal places. That is, the actual recorded values would be
$l_{T}=1.5$~[m], $l_{Sh}=1.681$~[m], $AE=1$~[m], and $BD=1.842$~[m].
Assuming the following measurement errors:
$\Delta l_{T}=0$~[m], $\Delta l_{Sh}=+0.005$~[m], $\Delta AE=+0.005$~[m], and $\Delta BD=-0.005$~[m],
we obtain the following erroneous measurements:
$l_{T}=1.5$~[m], $l_{Sh}=1.686$~[m], $AE=1.005$~[m], and $BD=1.837$~[m].
Using Eq.~\eqref{eqdelta}, we recover an angle $\bar{\delta}=87^{\circ}32'$. A measurement error of 5 millimeters thus leads to an error of about half a degree in the estimation of the angle. If this error is reduced to 1 millimeter, we obtain $\bar{\delta}=87^{\circ}5'$, corresponding to an error of 5 arcminutes in the angular measurement.\\

To perform this analysis more rigorously, we conducted a Monte Carlo simulation using data generated in Python, assuming that the measurement errors were normally distributed around the values rounded to three digits, and produced a set of 1000 measurements with errors. We show in TABLE~\ref{Tabla2}  the expected errors in the calculation of the angle as a function of the measurement errors in the distances from the ends of the tube to their projections and in the tube shadow length.

\begin{table}[h!]
\centering
\caption{\label{Tabla2}Expected errors in the calculation of the angle for $\alpha =30^{\circ}$ and $\delta=87^\circ$ as a function of the measurement errors, considering 1000 data points normally distributed around the exact value. Tube length 1.5~[m], outer diameter 1.1~[cm].}
\begin{ruledtabular}
\begin{tabular}{c c c }
 Measurement error (m) & Recovered mean & Standard deviation \\
\hline
$0.001$ & $86^\circ 58'$ & 5.38'   \\
$0.002$ & $86^\circ 58'$ & 10.97'  \\
$0.003$ & $86^\circ 58'$ & 16.05'  \\
$0.004$ & $86^\circ 58'$ & 22.17'  \\
$0.005$ & $86^\circ 58'$ & 27.52'  \\
\end{tabular}
\end{ruledtabular}
\end{table}

The simulation confirms that measurement errors range from about $5'$ for a 1~[mm] error, up to $27'$ in the case of 5~[mm] errors in the measurement of the shadow and the other distances involved in the calculation of the angle $\delta$ using Eq.~\eqref{eqdelta}.\\

The Python code used to perform these simulations is available in the repository
\texttt{https://github.com/Achenar-Star/Aristarco-Tube}.
By varying the value of $\alpha$, we observe that the errors are larger when the Moon is close to the horizon and decrease as it rises higher in the sky. In addition, increasing the tube length to 2 meters for the same angles further reduces the errors, which then range between $4'$ and $20'$, becoming even smaller when the Moon is higher in the sky.

\section{Choosing the time for the measurements}
\label{sec-time}

Any instructor wishing to replicate this experiment with students will encounter an immediate practical difficulty: even when the time of first or last quarter is known from lunar calendars~\cite{Fases}, the corresponding configuration may occur at an inconvenient time of day or under unfavorable observing conditions. Since the method requires the Sun and the Moon to be simultaneously visible, not every quarter phase is equally suitable for taking measurements. Although both objects may be visible on the same date, the interval of simultaneous visibility may occur several hours before or after the desired lunar phase.

The time of day at which the experiment can be performed is strongly constrained by the lunar phase. Near first quarter, the Moon is located approximately $90^\circ$ east of the Sun on the celestial sphere. It therefore rises roughly six hours after the Sun and remains visible during daylight mainly in the afternoon. Conversely, near last quarter the Moon is approximately $90^\circ$ west of the Sun, rising roughly six hours before the Sun and remaining simultaneously visible with it mainly during the morning. These six-hour intervals should only be regarded as approximate, since the actual rising and setting times also depend on the declinations of the Sun and the Moon.

The season of the year is also relevant. A useful qualitative picture is obtained by considering the motion of the Moon along the ecliptic. Near first quarter, the ecliptic longitude of the Moon is approximately $90^\circ$ ahead of that of the Sun, whereas near last quarter it is approximately $90^\circ$ behind. Consequently, at temperate latitudes the first-quarter Moon follows, approximately, the trajectory that the Sun follows one local season later, while the last-quarter Moon follows the trajectory of the Sun one local season earlier. This simple rule is independent of the hemisphere where the observer is located, and it
 is summarized in Table~\ref{tab:seasonal_observing}.

\begin{table}[h!]
\centering
\caption{Approximate seasonal behavior of the Moon near first and last quarter. The terminology refers to the trajectory followed by the Sun during the corresponding local season.}
\begin{ruledtabular}
\begin{tabular}{c c c}
Season & First quarter & Last quarter \\
\hline
Spring & summer-like & winter-like \\
Summer & equinox-like & equinox-like \\
Autumn & winter-like & summer-like \\
Winter & equinox-like & equinox-like \\
\end{tabular}
\end{ruledtabular}
\label{tab:seasonal_observing}
\end{table}

This observation gives a useful rule for planning the experiment. During spring, a first-quarter Moon follows a summer-like trajectory and can reach a relatively large altitude in the afternoon, whereas a last-quarter Moon follows a winter-like trajectory and may remain rather low in the morning. First quarter is therefore generally the more favorable configuration during spring. The situation is reversed during autumn: the last-quarter Moon follows a summer-like trajectory and is usually preferable to the first-quarter Moon, whose trajectory is winter-like. Around summer, both quarter Moons follow approximately equinox-like trajectories, and there is no strong geometrical preference for either configuration. During winter the Moon itself again follows an approximately equinox-like trajectory, but the Sun remains relatively low in the sky, making the experiment less favorable. The actual lunar altitude can of course depart from this simplified seasonal picture because the lunar orbit is inclined with respect to the ecliptic, so the predicted altitudes should ideally be checked beforehand with an astronomical ephemeris or planetarium program.

These considerations are particularly important for the present experiment because both objects should preferably be well above the horizon. As shown in Sec.~\ref{sec-accuracy}, the uncertainty in the recovered angle decreases when the Moon is higher in the sky. At the same time, a sufficiently high solar altitude produces a shorter and better-defined shadow, facilitating the accurate location of its endpoints. Observations with either object close to the horizon should therefore be avoided whenever possible.

To overcome the difficulty of observing exactly at the desired lunar phase, we measure the angle $\delta$ at successive time intervals several hours before and/or after first or last quarter and record the corresponding times and angular measurements. These observations should be made as close as possible to the desired configuration. The most favorable dates are those for which the Sun and Moon remain simultaneously observable at suitable altitudes for a sufficiently long interval on both sides of the quarter phase, allowing a large set of measurements to be obtained.

The resulting data are then used to estimate the value of $\delta$ at the desired time. When measurements are available both before and after the quarter phase, the angle is determined by linear interpolation. When observations can be obtained only before or only after it, the corresponding value is estimated by linear extrapolation. Interpolation is preferable whenever possible, since extrapolation becomes increasingly sensitive to the distance between the observation interval and the target time. In either case, combining several measurements reduces the influence of the uncertainty associated with any individual angular measurement and provides a more robust estimate of the desired angle.

The seasonal considerations described above provide useful general
guidelines, but they are not sufficient to determine whether the
experiment can actually be performed on a particular date. A more
restrictive condition is that, at the exact instant of lunar dichotomy,
both the Sun and the Moon must be simultaneously above the horizon.
Moreover, for practical purposes, it is highly desirable that neither
object be too close to the horizon.

In the following analysis we adopt an altitude of $15^\circ$ as a
practical lower limit. This value should not be interpreted as a sharp
physical threshold, but rather as a convenient criterion for selecting
favorable observing conditions. A low lunar altitude makes the
alignment of the tube more difficult, while a low solar altitude
produces longer and less sharply defined shadows. Therefore, events for
which both the Sun and the Moon are above approximately $15^\circ$ at
the instant of dichotomy are particularly suitable for the experiment.

As an illustrative example, we considered all waxing and waning lunar
dichotomies occurring during 2027 from the two cities in which the
authors of this work are based: Madrid, Spain, and Santiago, Chile.
Tables~\ref{tab:madrid2027} and~\ref{tab:santiago2027} list the only
events during that year for which both celestial bodies are above
$15^\circ$ at the instant when the lunar disk is exactly $50\%$
illuminated. The quoted times are local civil times.

\begin{table}[h!]
\centering
\caption{Favorable lunar dichotomies during 2027 as observed from
Madrid, Spain. Only events for which both the Sun and the Moon are
above $15^\circ$ altitude at the instant of dichotomy are included.
Times are local civil times.}
\label{tab:madrid2027}
\begin{ruledtabular}
\begin{tabular}{lcccc}
Date & Phase & Time & $h_{\odot}$ & $h_{\rm M}$ \\
\hline
15 Mar. & Waxing  & 16:10 & $33.1^\circ$ & $49.2^\circ$ \\
10 Jul. & Waxing  & 19:42 & $21.2^\circ$ & $37.2^\circ$ \\
23 Sep. & Waning  & 14:00 & $49.4^\circ$ & $18.1^\circ$ \\
20 Dec. & Waning  & 11:04 & $19.5^\circ$ & $19.8^\circ$ \\
\end{tabular}
\end{ruledtabular}
\end{table}

\begin{table}[h!]
\centering
\caption{Favorable lunar dichotomies during 2027 as observed from
Santiago, Chile. Only events for which both the Sun and the Moon are
above $15^\circ$ altitude at the instant of dichotomy are included.
Times are local civil times.}
\label{tab:santiago2027}
\begin{ruledtabular}
\begin{tabular}{lcccc}
Date & Phase & Time & $h_{\odot}$ & $h_{\rm M}$ \\
\hline
15 Jan. & Waxing  & 15:36 & $63.9^\circ$ & $16.8^\circ$ \\
29 Jan. & Waning  & 08:59 & $22.8^\circ$ & $65.7^\circ$ \\
28 May  & Waning  & 11:20 & $31.9^\circ$ & $22.2^\circ$ \\
10 Jul. & Waxing  & 13:20 & $33.9^\circ$ & $16.4^\circ$ \\
7 Sep.  & Waxing  & 13:42 & $50.6^\circ$ & $16.9^\circ$ \\
\end{tabular}
\end{ruledtabular}
\end{table}

These results illustrate that the seasonal rules discussed above are
only a first step in selecting an observing date. For example, the
waxing dichotomy of March 15 is particularly favorable from Madrid:
at the exact instant of dichotomy the Sun and Moon reach altitudes of
approximately $33^\circ$ and $49^\circ$, respectively. This agrees
with the general expectation that a waxing Moon is preferable during
spring at northern mid-latitudes.

Similarly, from Santiago the waning dichotomies of January 29 and
May 28 provide particularly favorable configurations. The latter is
especially consistent with the seasonal argument above, since May
corresponds to autumn in the Southern Hemisphere, when a waning Moon
follows a comparatively high trajectory in the morning sky.

These examples also emphasize that the optimum observing dates must be
determined for the particular year and geographical location of the
experiment. The seasonal considerations provide useful qualitative
guidance, whereas the actual solar and lunar altitudes at the instant
of dichotomy ultimately determine whether a given event is suitable.

For secondary-school students, the instructor may provide in advance
the observing window and the required astronomical ephemerides, allowing
the students to concentrate on the construction of the instrument, the
acquisition of the measurements, and the geometrical determination of
the Sun--Earth--Moon angle.

\section{Real data and estimation of the angle}
\label{sec-realdata}

We tested our instrument with excellent results and provide several recommendations to improve the accuracy of the measurements. First, it is highly desirable to work in teams of two to five people. One person should be responsible for aiming at the Moon by looking through the interior of the tube and adjusting the tripod controls so that the tube remains fixed during the measurement. Immediately afterward, two markers must be placed at the ends of the tube’s shadow on the ground, taking care to position them accurately. This step is essential because the shadow of the tube moves rapidly and can change noticeably within about thirty seconds.

Once the ends of the shadow are marked, the three required distances are measured: from the lower end of the tube to the first marker, from the upper end of the tube to the second marker, and the distance between the two markers. Care must be taken not to move the markers during this process, as doing so would invalidate the measurement. The time at which the Moon was first aligned through the tube, together with all measured distances, should be recorded in a notebook. Ideally, the ground surface should be smooth, and the lower end of the tube should be kept as close to the ground as possible, even if this requires the observer to lie on the ground in order to properly aim at the Moon.

\begin{table}[h!]
\centering
\caption{Measurements taken on October 10, 2024, from Santiago, Chile. Tube length $l_T=1.5$~[m]. The angle $\delta$ is calculated using Eq.~\eqref{eqdelta}.The Exact reference values are obtained from the Stellarium software and the error of the calculated angle with respect to exact value.}
\begin{ruledtabular}
\begin{tabular}{c c c c c c c} 
 Time & AE & BD & $l_{Sh}$ & $\delta$ & Exact & Error \\
\hline
15:00 & $0.855$ & $1.752$ & $1.765$  & $89^{\circ}49'$ & $90^{\circ}19'$ & $-30'$ \\
15:08 & $0.85$ & $1.787$ & $1.81$  & $90^{\circ}25'$ & $90^{\circ}23'$ & $+2'$ \\
15:16 & $0.84$ & $1.816$ & $1.865$  & $89^{\circ}56'$ & $90^{\circ}28'$ & $-32'$ \\
15:19 & $0.84$ & $1.836$ & $1.885$ & $90^{\circ}31'$ & $90^{\circ}27'$ & $+4'$ \\
15:24 & $0.83$ & $1.857$ & $1.916$ & $90^{\circ}20'$ & $90^{\circ}30'$ & $-10'$ \\
15:27 & $0.827$ & $1.869$ & $1.935$ & $90^{\circ}16'$ & $90^{\circ}31'$ & $-15'$ \\
15:31 & $0.822$ & $1.89$ & $1.96$  & $90^{\circ}27'$ & $90^{\circ}32'$ & $-5'$ \\
15:43 & $0.815$ & $1.946$ & $2.044$  & $90^{\circ}24'$ & $90^{\circ}37'$ & $-13'$ \\
15:48 & $0.811$ & $1.977$ & $2.085$ & $90^{\circ}31'$ & $90^{\circ}39'$ & $-8'$ \\
15:53 & $0.806$ & $2.015$ & $2.135$ & $90^{\circ}37'$ & $90^{\circ}41'$ & $-5'$ \\
15:59 & $0.803$ & $2.046$ & $2.183$ & $90^{\circ}26'$ & $90^{\circ}44'$ & $-18'$ \\
16:02 & $0.805$ & $2.065$ & $2.204$ & $90^{\circ}46'$ & $90^{\circ}45'$ & $+1'$ \\
16:08 & $0.805$ & $2.121$ & $2.277$ & $91^{\circ}03'$ & $90^{\circ}47'$ & $+16'$ \\
16:15 & $0.925$ & $2.144$ & $2.442$ & $90^{\circ}34'$ & $90^{\circ}50'$ & $-16'$ \\
16:20 & $0.847$ & $2.148$ & $2.368$ & $90^{\circ}38'$ & $90^{\circ}51'$ & $-13'$ \\
16:25 & $0.85$ & $2.183$ & $2.41$  & $91^{\circ}0'$  & $90^{\circ}53'$ & $+7'$ \\
16:30 & $0.85$ & $2.228$ & $2.468$  & $91^{\circ}8'$  & $90^{\circ}55'$ & $+13'$ \\
\end{tabular}
\end{ruledtabular}
\label{Tabla1Tubo}
\end{table}

On October 10, 2024, the last quarter occurred at 13:57 local time in Santiago, Chile. Our goal was to identify a site within the city with a clear eastern horizon and minimal obstruction from the Andes mountain range. The selected location was the Foster Observatory on San Crist\'obal Hill \cite{Foster}, which offers a flat terrace and an excellent view of the eastern horizon over the city. The Moon was expected to rise at approximately 12:30. Our initial plan was to obtain multiple measurements both before and after the moment of last quarter, including a measurement as close as possible to the exact time of the quarter.

As often occurs in astronomical observations, weather conditions interfered with the planned measurements. A dense layer of clouds formed along the eastern horizon, delaying the observations, despite otherwise clear skies and full sunlight producing a well-defined shadow of the tube on the ground. The first usable data were obtained after 15:00, once the Moon had risen above the cloud layer and both the Sun and the Moon were simultaneously visible.

With careful coordination, one observer aligned the Moon through the tube. Once the Moon was clearly visible and centered inside the tube, two markers were placed to define the ends of the shadow. For this purpose, we used small wooden sticks similar to those commonly employed as coffee stirrers. This step must be performed quickly, as the shadow moves rapidly, and steady coordination is required to measure accurately the distances from the ends of the tube to their projections on the ground, as well as the length of the shadow itself. Two observers performed the measurements, while a third recorded the time and the measured values. All measurements were taken using a standard measuring tape with millimeter resolution, readily available at any hardware store. The data obtained that day are summarized in Table~\ref{Tabla1Tubo}.

\begin{table}[h!]
\centering
\begin{ruledtabular}
\begin{tabular}{c c c}
\textbf{Time (min)} & \textbf{Time (h)} & \textbf{Angle $\delta$ (degrees)} \\
\hline
0  & 0.000 & 89.81959150 \\
8  & 0.133 & 90.43308460 \\
16 & 0.267 & 89.94840580 \\
19 & 0.317 & 90.52765860 \\
24 & 0.400 & 90.33507140 \\
27 & 0.450 & 90.26712270 \\
31 & 0.517 & 90.45407460 \\
43 & 0.717 & 90.41142360 \\
48 & 0.800 & 90.53148520 \\
53 & 0.883 & 90.63209150 \\
59 & 0.983 & 90.43893980 \\
62 & 1.033 & 90.77849300 \\
68 & 1.133 & 91.06212170 \\
75 & 1.250 & 90.57217420 \\
80 & 1.333 & 90.63365680 \\
85 & 1.417 & 91.00259260 \\
90 & 1.500 & 91.13394970 \\
\end{tabular}
\end{ruledtabular}
\caption{\label{Tabla4}Data used for the linear fit to estimate the angle $\delta$ at last quarter.}
\end{table}

\begin{figure}[h!]
\centering
\includegraphics[width=8cm]{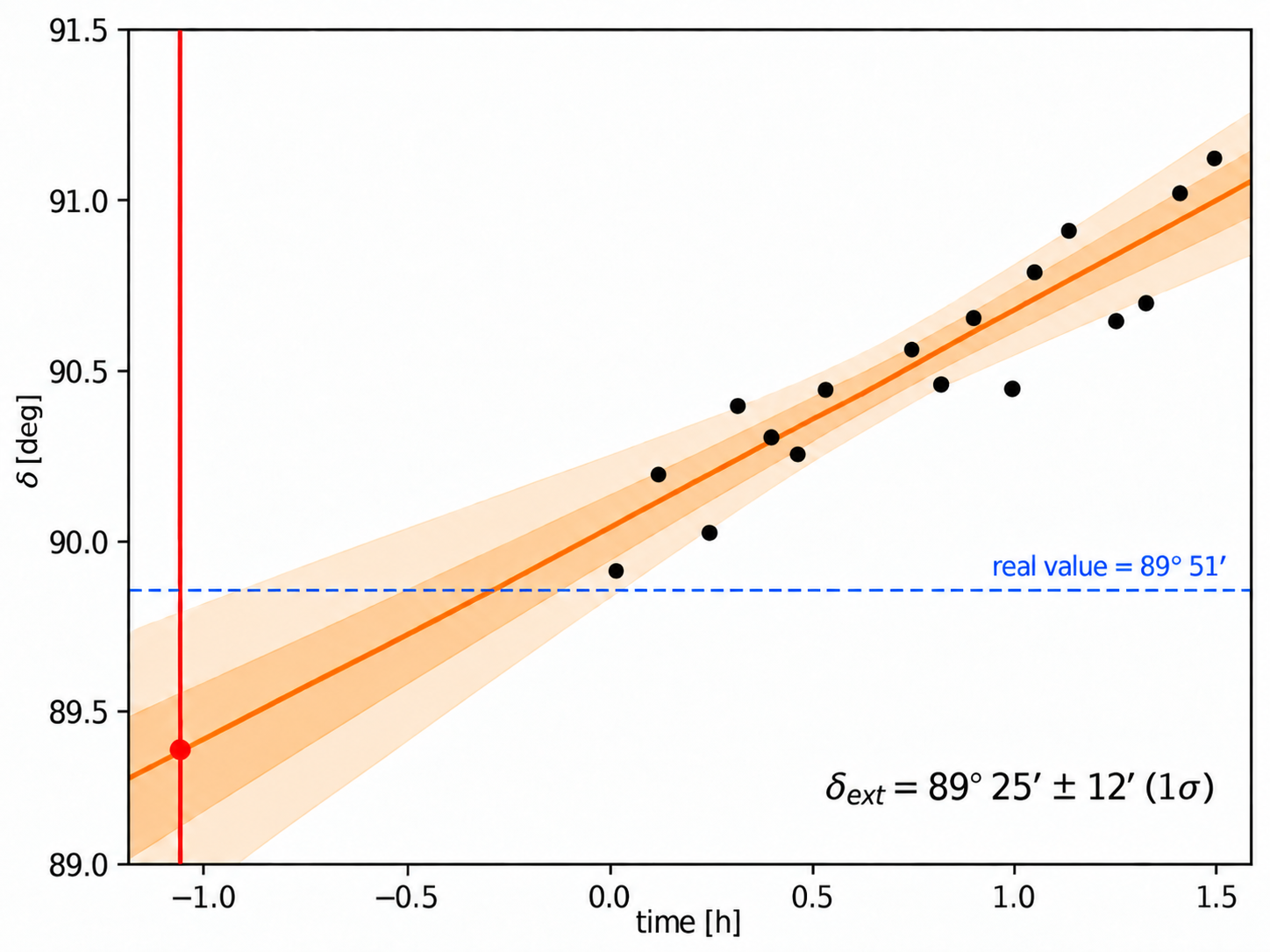}
\caption{Correlation plot and linear fit for the data obtained on October 10, 2024. Experimental values of $\delta$ and their linear fit. The shaded region shows the regression uncertainty. Extrapolation gives $\delta_{\mathrm{ext}} = 89^\circ 25' \pm 12'$ ($1\sigma$), while the dashed line indicates the reference value $\delta = 89^\circ 51'$.}
\label{LinError}
\end{figure}

Since the last quarter occurred approximately one hour before our first measurement, we performed a linear fit to the measured data in Table~\ref{Tabla4}, as we see in Fig.~\ref{LinError}  and obtained
$$
\delta(t)=90.05455+0.61329\,t,
$$
Since the exact time of the first quarter moon—according to Stellarium—is 13:57 on that day, and considering 15:00 as our time origin (in minutes), this corresponds to $-63$ min in this system. Extrapolating using the obtained line. That means to evaluate  this expression at $t=-1.05$ hours, where the chosen time origin corresponds to 15:00, yields our final estimate for the angle:
\begin{center}
$89.410596^\circ$, that is, $\delta\approx 89^\circ 25'$.
\end{center}
$$\lambda\approx 97.21$$

The uncertainty associated with the extrapolated angle was estimated from the linear regression of the experimental data \cite{Montgomery21}. A 68\% confidence interval was calculated using the residual standard error of the fit and the distance between the extrapolated time and the mean of the measured times. Since the extrapolation was performed outside the experimental time range, the associated uncertainty increases with the distance from the measured data.
$$\delta = 89^\circ 25' \pm 12'$$
Because $t_0$ lies outside the measured time interval, the uncertainty increases as the extrapolation point moves farther from the mean of the experimental data as we can see in FIG.~\ref{LinError}.

\section{Description of the errors associated with the angle measurement}
\label{sec-errors}

\begin{figure}[h!]
\centering
\includegraphics[width=7cm]{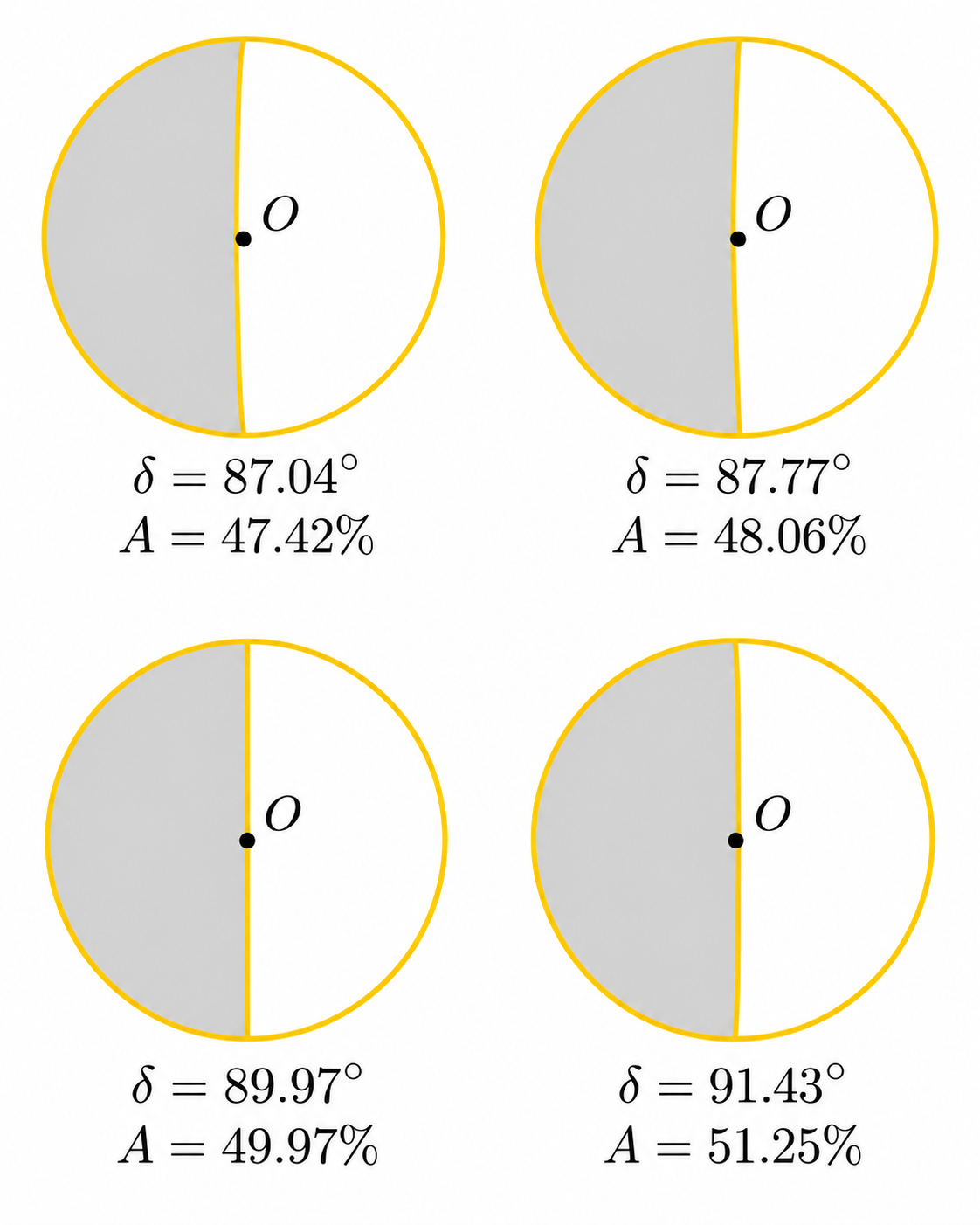}
\caption{The figure shows the elliptical curve drawn on the Moon at the time of Aristarchus’ measurement for an angle of $87^\circ$, as well as at moments before, during, and after first quarter. Inferring the exact moment of first quarter without instruments from the shape of the shadow is extremely difficult, particularly because of the low visibility during daylight and the irregularities of the shadow near the lunar terminator.}
\label{Som}
\end{figure}

First, the dominant source of error identified in this experiment is the difficulty in determining the exact moment of first or last quarter. It should be noted that Aristarchus had access only to the shadow cast on the Moon to determine the moment of first quarter, without instruments. Even with small telescopes or binoculars, distinguishing the precise shape of the terminator curve during daylight is quite challenging. Figure~\ref{Som} shows a model of the shadow at times shortly before, during, and after first quarter, constructed using the GeoGebra animation previously presented in Ref.~\cite{ConMario}, where we studied the shape of the lunar terminator. The Moon appears very distant, and over several minutes the shaded line seems essentially unchanged; it is therefore almost impossible to detect differences in the shadow over the course of an hour, during which the angle changes by approximately half a degree.\\

Second, there is an error associated with aiming at the Moon through the interior of the tube. If the tube diameter is smaller than the one indicated in our design, an additional error is introduced because it becomes difficult to determine whether the Moon is precisely centered with respect to the tube. This misalignment can lead to errors of several arcminutes in the final angle, depending on how well centered the Moon is. It is important to recall that the Moon is observed while it is half illuminated and during daylight, which significantly reduces its visibility.\\

Third, there is an error associated with identifying the ends of the tube’s shadow on the ground. When the Sun is high in the sky, these endpoints are well defined. However, as the Sun lowers, the shadow length increases and diffraction effects appear at the upper end of the tube. At that point, it becomes difficult to place the marker accurately at that end of the shadow, and we recommend discontinuing the measurements, as the errors begin to increase significantly.\\

A fourth source of error arises when taking the measurements with the measuring tape. In some cases, it is difficult to place one end of the tape on the ground near the first marker, and since the measurement is taken to the center of the tube, which is not physically marked, additional uncertainties are introduced. These errors have already been analyzed using the model presented in Sec.~\ref{sec-accuracy}.\\

\section{Discussion and conclusions}
\label{sec-conclusions}

Throughout this work we have revisited, from a modern perspective, the problem of determining the ratio between the Earth–Sun and Earth–Moon distances, in effect recreating the famous experiment of Aristarchus. Those willing to carry out this experience will gain a genuine appreciation for the value of Aristarchus’ original estimate. The value of $87^\circ$ for the Sun–Earth–Moon angle at the time of first quarter is undoubtedly the result of considerable effort, careful observation of the motions of the Sun and the Moon, and remarkable creativity in recognizing that under these conditions the desired ratio of distances could be determined.\\

We presented a low-cost instrument specifically designed to measure this angle, which allows trigonometry to be introduced in a practical way with a concrete astronomical goal in mind. The procedure connects angles and linear measurements to infer quantities that may seem inaccessible to those not directly engaged in the study of science.\\

The model developed in this work allows us to analyze the expected errors in the angle measurement by varying parameters such as the length of the tube and the altitude of the Moon at the time of observation. The experimental results obtained are consistent with the predictions of the model. The ability to test the expected precision of the instrument prior to its construction, and to influence its design in order to achieve optimal performance, constitutes one of the main strengths of the proposed method. This approach also introduces desirable elements of programming and modeling, which are valuable in any activity aimed at developing scientific skills in students.\\

\begin{figure}[h!]
\centering
\includegraphics[width=8cm]{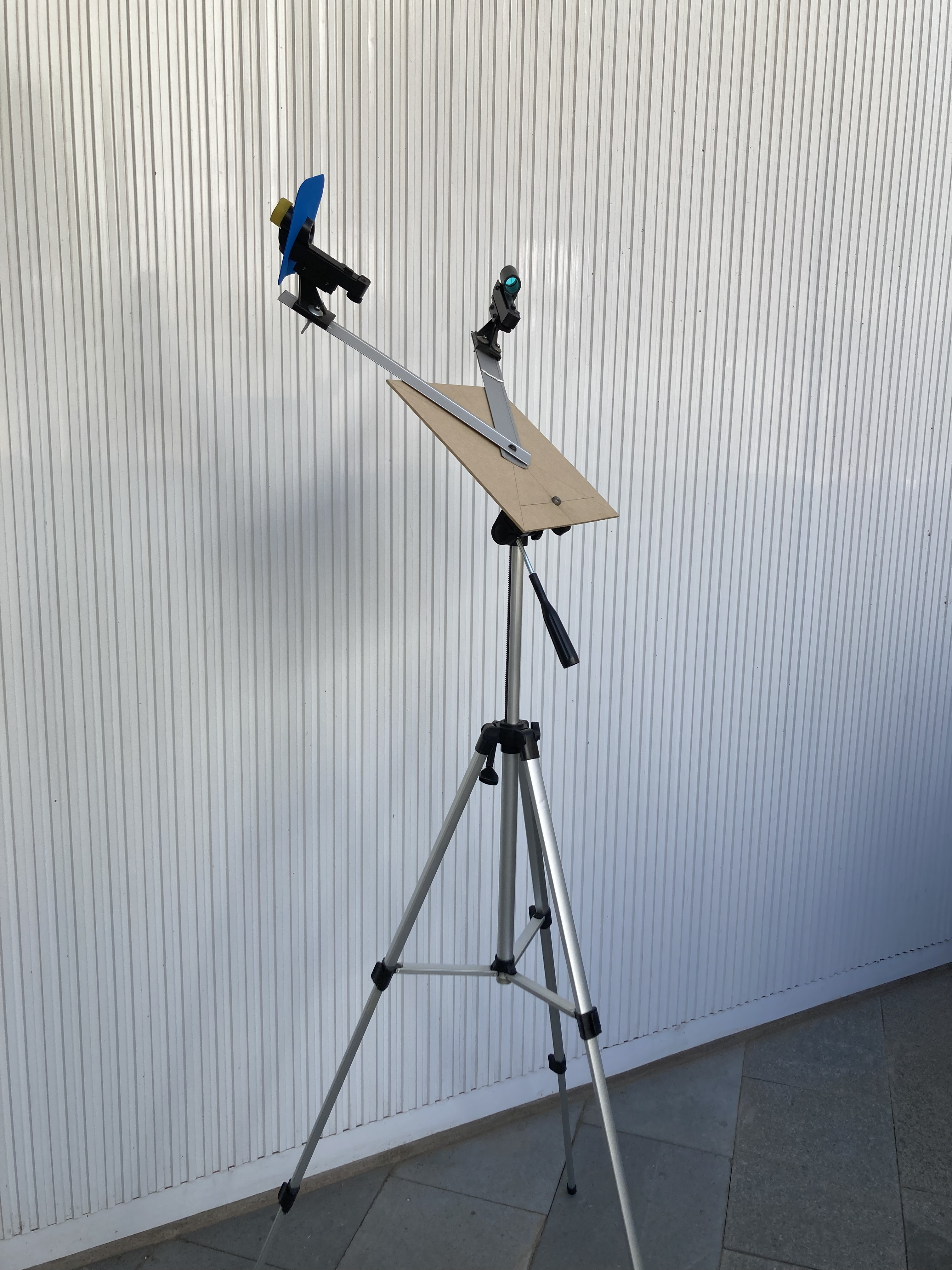}
\caption{The figure shows the first instrument constructed. One pointer is used to aim at the Moon and the other at the Sun. A bottle cap fitted with a solar filter and a foam rubber shield are used to allow safe pointing toward the Sun without glare.}
\label{Diamante}
\end{figure}

An important issue to highlight is that our initial investigation of this problem, led us to construct a direct-measurement instrument that included astronomical laser pointers, with the angle measured using a digital protractor. However, we encountered significant difficulties when attempting to simultaneously aim at both the Sun and the Moon. Because the apparent diameters of the Sun and the Moon are approximately half a degree, a misalignment of even one degree in the mounting platform causes both objects to fall out of view. As a result, this direct-measurement approach exhibited multiple sources of error and never produced results of the same quality as those obtained with the simple instrument presented in this work. Table~\ref{Tabla Lasers} summarizes the best results obtained using the digital protractor while simultaneously aiming at the Sun and the Moon, and Fig.~\ref{Diamante} shows the instrument used to obtain these data.

\begin{table}[h!]
\centering
\caption{Measured values of the angle $\delta$ on the day of first quarter, July 13, 2024, from Santiago, Chile. Angles smaller than $90^{\circ}$ obtained using a direct-measurement instrument with a digital protractor.}
\begin{ruledtabular}
\begin{tabular}{c c c c r}
 Time & $\delta$ & Exact & Error & $\sec(\delta)=\lambda$ \\
\hline
14:27 & $88^{\circ}12'$  & $88^{\circ} 27'$ &+15& 31.8 \\
16:09 & $88^{\circ}57'$ & $89^{\circ} 02'$ &-5 & 54.6 \\

\end{tabular}
\end{ruledtabular}
\label{Tabla Lasers}
\end{table}

Here the main problem is the difficulty of obtaining measurements—which are prone to multiple errors involving the alignment of the sun and moon with the instrument—should prompt the experimenter to question whether a direct measurement can truly be obtained and validated as reliable data; therefore, we unreservedly recommend the measurement method using the tube, which undoubtedly yields results we consider more trustworthy.

One way to further reduce the random uncertainty is to perform the
experiment simultaneously at multiple locations. Depending on whether
the time of first or last quarter falls within or outside the observation
interval, interpolation or extrapolation may be used. If each site
provides an independent estimate with a standard uncertainty of
\(u_{\mathrm{ind}}=0.5^\circ=30'\), as might be expected when the
measurements are performed by students, the standard uncertainty of the
mean of \(n\) independent measurements is given by
\begin{equation}
    u_{\mathrm{mean}}
    = \frac{u_{\mathrm{ind}}}{\sqrt{n}},
    \label{eq:uncertainty_mean}
\end{equation}
as follows from the standard statistical treatment of independent random
uncertainties \cite[Chap.~4]{Taylor97}. Therefore, requiring
\(u_{\mathrm{mean}}=2'\) gives
\begin{equation}
    \frac{30'}{\sqrt{n}}=2',
    \qquad
    n=\left(\frac{30'}{2'}\right)^2=225.
\end{equation}
Thus, under the assumption that the individual estimates are
statistically independent and affected predominantly by random errors,
225 simultaneous measurements would reduce the standard uncertainty of
their mean to \(2'\). Fewer measurements would be required if the
individual estimates were more precise. For example, if
\(u_{\mathrm{ind}}=10'\), then
\begin{equation}
    n=\left(\frac{10'}{2'}\right)^2=25.
\end{equation}
The experiment therefore has an inherently collaborative component:
measurements obtained at different locations can be combined to improve
the precision of the final estimate.\\

Discussions with several instructors during previous presentations of this work have yielded positive feedback regarding the pointer-based instrument, as it—or modified versions of it—can be highly effective for teaching trigonometry while providing a practical component that is often missing in traditional instructional approaches. This aspect is particularly important for promoting meaningful learning of essential tools such as trigonometry and geometry in scientific education.\\

More broadly, the experiment provides a versatile scientific and
inquiry-based activity that can be adapted to students at different
educational levels. At secondary-school level, the emphasis may be
placed on the construction of the instrument, experimental
measurements, geometry, and trigonometry. At university level, the
same activity can be extended to include astronomical ephemerides,
error propagation, statistical analysis, linear regression, numerical
modeling, and programming. The instructor can therefore adjust the
mathematical, computational, and experimental depth of the activity
without changing its underlying scientific question.

Figure~\ref{Maza} also suggests an alternative approach for determining the angle. This method involves taking photographs of the Moon before and after first and last quarter, counting pixels, and interpolating to determine the exact time of each quarter phase. The resulting time difference, as discussed in the introduction, should be much smaller than one day and can be used to estimate the angle $\delta$. Although we have not yet implemented this approach, we are actively developing these ideas, which will allow us to present a novel method for addressing this problem. The tools involved are of particular interest at the undergraduate level, as they incorporate more advanced technological and mathematical elements suitable for higher-level courses.\\

The instrument presented in this work has the advantage of being very easy to construct, making it suitable for use in science fairs and outreach activities that promote the integration of historical, astronomical, physical, geometrical, and computational elements. Together, these aspects make this experience a unique opportunity for those willing to follow in the footsteps of Aristarchus in estimating the distance to our favorite star, the Sun.\\

\begin{acknowledgments}
The authors thank the Foster Observatory for allowing us to collect the main data at this site, and Dami\'an Pacheco of the Foster Observatory for his assistance during the data acquisition. H.~C.\ acknowledges the Grupo de Observaci\'on Astron\'omica of Universidad Adolfo Ib\'a\~nez for their support. 
The authors acknowledge the use of artificial intelligence tools to improve the visual presentation of some figures.
\end{acknowledgments}


\end{document}